\documentclass[aps,twocolumn,showpacs]{revtex4b5}

\def\mnras{Mon. Not. R. Astr. Soc. }

\usepackage{graphicx}
\usepackage{dcolumn}
\usepackage{amsmath}

\begin{document}
\title{A New Connection between Central Engine Weak Physics and the Dynamics 
of Gamma-Ray Burst Fireballs}

\author{Jason Pruet}
\email{jpruet@physics.ucsd.edu}
\author{Kevork Abazajian}
\email{kabazajian@ucsd.edu}
\author{George M. Fuller}
\email{gfuller@ucsd.edu}
\affiliation{Department of Physics, University of California, San Diego,
La Jolla, California 92093-0319}
\date{January 10, 2001}

\begin{abstract}
We demonstrate a qualitatively new aspect of the dynamics of Gamma-Ray
Burst (GRB) fireballs: the development of a substantial dispersion in
the proton component in fireballs in which neutron decoupling occurs
and is sufficiently pronounced.  This effect depends sensitively on
the neutron to proton ratio in the fireball, becoming more dramatic
with increasing neutron excess.  Simple physical arguments and
transport calculations indicate that the dispersion in Lorentz factor
of the protons can be of order the final mean Lorentz factor of the
fireball. We show how plasma instabilities could play an important
role in the evolution of the fireball and how they might ultimately
govern the development of such a velocity dispersion in the proton
component.  The role of these instabilities in setting/diminishing a
proton Lorentz factor dispersion represents a new and potentially
important venue for the study of plasma instabilities.  Significant
dispersion in the proton velocities translates into {\it fewer}
protons attaining the highest Lorentz factors. This is tantamount to a
reduction in the total energy required to attain a given Lorentz
factor for the highest energy protons. As well, a proton component
dispersion can have consequences for the electromagnetic and neutrino
signature of GRBs.
\end{abstract}

\pacs{98.70.Rz, 14.60.Pq}
\maketitle

\section{Introduction}
We show that a simple, yet previously overlooked, mechanism may give
rise to a substantial velocity dispersion in the proton component of
Gamma-Ray Burst fireballs. Namely, in GRB fireballs in which neutron
decoupling occurs, large center of mass energy collisions between
decoupled neutrons and protons will induce a velocity dispersion in
the proton component. For certain fireball parameters the timescale
for re-thermalization of these protons is longer than the dynamical
timescale, and so the protons will retain their acquired dispersion.

This has consequences for the dynamics of the fireball.  Since the
ultra-relativistic proton shell is directly tied to gamma-ray
photon production \cite{rm,sp}, the distribution of proton energy
can affect the dynamics of photon emission.  In the internal shock 
GRB model
the $\gamma$-ray photons are produced through internal multiple-shock
collisions involving the protons. The efficiency for the conversion
of shock kinetic energy to photon energy increases with increasing
relative shock velocities \cite{rm,piran}. In addition, the energetic proton
shock produces particle cascades leading to multi-TeV neutrino
emission \cite{waxbah,halzentasi,halzenhooper}. Therefore, both
photon and neutrino radiation may be affected by the process discussed
here.

More basically, the process we discuss here represents 
a new aspect of our understanding of the evolution of homogeneous
relativistic fireballs. The standard picture 
holds that a thermal plasma undergoes
an initial acceleration phase, lasting until the energy in
relativistic particles is transferred to the kinetic energy of protons,
or, in the case of extremely baryon-dilute plasmas, until the plasma
becomes optically thin to photons \cite{shp90}. Following the
acceleration phase a coasting phase ensues. In the coasting phase the
baryons are assumed thermal with typically low temperatures,
$T<10^{-4}\ {\rm MeV}$, in the plasma rest frame.

Recently it has been recognized that in fireballs satisfying certain
initial conditions the neutrons will dynamically decouple 
\cite{stellar,derishev,fpa} from the
expanding $e^{\pm}$/photon/proton plasma. There is an interesting connection
between this decoupling and the net number of electrons per baryon,
\begin{equation}
Y_e=\frac{n_{e^-}-n_{e^+}}{n_b}=\frac{1}{n/p+1},
\end{equation} 
where $Y_e$ is the electron fraction, 
$n_{e^+}$ and $n_{e^-}$ are the number densities of positrons and 
electrons respectively, and $n$ and $p$ are the number densities of neutrons
and protons.
The connection is that when  conditions near the
fireball source lead to a low $Y_e$ in
the outflow, substantial differences between the final neutron and
proton Lorentz factors are possible \cite{fpa}. 
We will demonstrate a second connection:
when the electron fraction in the fireball is low, 
 high energy collisions
between decoupled neutrons and protons induce a velocity dispersion
in the protons and that   
these ``hot protons'' are not re-thermalized. The sensitivity of this
effect to $Y_e$ in the fireball is interesting because, in analogy to 
type II supernovae, the electron fraction may mirror weak physics
in the central engine \cite{qian,fpa}.

\section{Neutron Decoupling in relativistic fireballs}
The essential features of the physics of baryon flow in relativistic
fireballs are obtained by considering a two component [(i)
$e^\pm$/photon/proton and (ii) neutron] homogeneous fireball with
initial radius,
temperature, and electron fraction $R_0,T_0,$ and $Y_{e0}$
respectively. In terms of these quantities the entropy per baryon in 
units of $10^5 k_{\rm b}$ is 
\begin{equation}
s_5 \approx 1.250 \times
10^{-2}\ \eta (1\ {\rm MeV}/T_0).
\end{equation}
Here $\eta=E_{\rm tot}/M$ is  the ratio of total 
energy to rest mass in the fireball. A central issue for GRBs is the 
``baryon loading problem'', the statement that $\eta$ must be large
in order to get the fireball moving with a high Lorentz factor.

For relativistic fireballs ($\eta >$ a few), numerical and
analytic work show that to first approximation the fireball evolves
as  \cite{kobayashiandpiran,shemiandpiran}
\begin{eqnarray}
\gamma= (T_0/T)=R/R_0 &{\rm for }& \gamma < \eta \\
\gamma=\eta \,\,\,\,\,\,\,\,\,\,\, &{\rm for}& R > \eta R_0.
\end{eqnarray} Here $\gamma$
is the Lorentz factor of the fireball. These relations follow from
entropy and energy conservation and are violated at the beginning and
end of the fireball's evolution (see below).  In terms of the time $t$
as measured by an observer comoving with the plasma, the Lorentz
factor and temperature evolve as $\gamma=(T_0/T)= e^{t/\tau_{\rm
dyn}}$. Here $\tau_{\rm dyn}=R_0$ characterizes the fireball source
size. (We adopt units where the speed of light is
unity.) Observations of time variability in GRBs give the 
constraint $\tau_{\rm dyn}<10^{-3}\rm\ s$, while the smallest proposed
GRB sources are solar mass scale compact objects with $\tau_{\rm dyn}
\ge 10^{-5}\rm\ s$.

Particles in the plasma 
suffer a 4-acceleration  with magnitude gamma
$d\gamma/dR \approx R_0^{-1}$. Neutrons have such a
small magnetic dipole moment that they are essentially only coupled to
the accelerating plasma via strong scatterings with protons (near the
decoupling point $\sigma_{\rm np} \sim 0.1 \sigma_{\rm T} \sim 10^7
\sigma_{{\rm n}e} \sim 10^{15} \sigma_{{\rm n}\gamma}$) where $\sigma_{\rm T}$ is
the Thomson cross section. To an excellent approximation, then, the
neutron-proton collision timescale in terms of the comoving proton
number density $n_{\rm p}$ in the plasma rest frame is 
\begin{equation}
\tau_{\rm
coll}^{-1}=n_{\rm p} \sigma_{\rm np} v_{\rm rel},
\end{equation}
 where $v_{\rm rel}$ is the
relative neutron proton velocity.  Apart from small corrections
stemming from inelastic nucleon-nucleon scatterings, entropy is
conserved throughout the acceleration of the fireball. This implies
 \begin{equation}      \tau_{\rm coll}^{-1} \propto n_{\rm p} \propto T^3 \propto
\exp{-3t/\tau_{\rm dyn}}\end{equation} except for the brief period when the
electron/positron pairs annihilate and transfer their entropy to the
photons.

Decoupling occurs if the dynamic
and collision 
timescales become comparable before the end of the acceleration phase
of the fireball's evolution. This condition is
\begin{equation}
\label{decoupling}
{\frac{0.02\ s_5^4}{\tau_{5} Y_e [v_{\rm rel}\sigma_{10}]_{\rm dec}}} > 1. 
\end{equation}
Here $\tau_{5}$ is the dynamic timescale in units of
$10^{-5}\ \rm s$, and $(v_{\rm rel} \sigma_{10})_{\rm dec}$ is the
product of the relative velocity and neutron/proton cross section in
units of 10 ${\rm fm}^2$ at the decoupling point. 
The precise $v_{\rm rel}$ at which this
latter quantity should be evaluated is unclear from our simple
argument. This is not crucial because $[v_{\rm rel} \sigma_{10}]$
only varies by a factor of 5 from unity as $v_{\rm rel}$
increases from $10^{-3}$ to $1$.  The final Lorentz factor of the neutrons is
estimated by evaluating $\gamma$ at the decoupling point, 
\begin{equation}
\gamma_{\rm n,\rm final}\approx 220 \left(\frac{[v_{\rm rel}
\sigma_{10}]_{\rm dec} T_{0,{\rm MeV}}^3 Y_e \tau_{5}}{s_5}
\right)^{1/3}.
\end{equation}

Following decoupling, the proton/$e^{\pm}$/photon plasma begins
to accelerate away from the neutrons. Energy conservation gives an
estimate of the final mean Lorentz factor of the protons,
\begin{equation}
\label{gammabarp}
\langle{\gamma_{\rm p,{\rm final}}}\rangle
\approx{Y_e^{-1}} \left[\eta-\gamma_{\rm n,final}
 \left(1-Y_e+ \epsilon
{\frac{E_{\nu}}{m_n}\frac{\gamma_{\pi}}{\gamma_{\rm n,\rm final}}}\right)
\right].
\end{equation}

The term $\epsilon (E_{\nu}/m_n) (\gamma_{\pi} /\gamma_{\rm n,final})$ 
is present
to account for energy lost to neutrinos arising from the decay of
pions created in inelastic nucleon-nucleon collissions. Here
$m_n$ is the nucleon mass, $\epsilon$ is the number of pions created
per baryon (at most of order unity), $E_{\nu}$ is the average center
of mass energy lost to neutrinos per pion decay, and $\gamma_{\pi}$ is
the average Lorentz factor of created pions. For pions created in n-p
collisions, the branching ratio for charged pion production below the
two pion threshold is roughly 1/2, and $E_{\nu} \sim 50 {\rm\,
MeV}$. It is difficult to analytically estimate $\gamma_{\pi}$, but as
most inelastic nucleon-nucleon collisions occur near the decoupling
point, it is presumably of order $\gamma_{\rm n,final}$.  
In the numerical examples
we present below, neutrino energy loss is estimated 
and is found to be small enough so as not to alter the qualitative
picture presented here.

\section{Neutron induced proton heating in relativistic fireballs}

Subsequent to neutron decoupling, protons will undergo collisions with
decoupled neutrons. 
The first requirement for the presence of a ``hot'' proton component is
that some of these collisions are high energy collisions.  
If in the lab frame a proton and neutron have Lorentz factors
$\gamma_{\rm p}$ and $\gamma_{\rm n}$ respectively, then the $\gamma$ of the
neutron as seen by the proton is 
\begin{equation}
\gamma_{\rm rel}=\frac{1}{2} 
\frac{\gamma_{\rm p}}{\gamma_{\rm n}}
\left(1+(\gamma_{\rm n}/\gamma_{\rm p})^2\right).
\end{equation}
 A high energy 
neutron-proton collision implies 
then:
$ 
{\langle{\gamma_{\rm p,{\rm final}}}\rangle}/{\gamma_{\rm n,\rm final}} \ge {\rm a\,few}.
$
This condition is intendend only as a
rough guide for the fireball parameters needed to drive the proton
component hot. 

For reasonable fireball parameters it is difficult to
attain substantial 
$ {\langle{\gamma_{\rm p,{\rm final}}}\rangle}/{\gamma_{\rm n,\rm final}} $ 
unless $Y_e < 0.5$. It has been argued
that a low electron fraction will naturally be obtained in many of the
proposed GRB environments \cite{fpa}. In addition, in calculations of
GRBs arising from neutron star mergers  $Y_e\approx 0.1$
has been estimated \cite{salmwilson}. It is not clear if such low electron
fractions are also obtained, for example, in the collapsar model for GRB
central engines \citep{woosley}.

Protons undergoing collisions with neutrons 
will lose energy to the background photons
and electrons. Whether or not these
protons re-thermalize with the plasma depends on their thermalization
timescale $\tau_{\rm therm}$.  The protons will remain
``cool'' ({\it i.e.,} well coupled and thermal) 
if $\tau_{\rm therm} < \tau_{\rm dyn}( \le
\tau_{\rm coll})$, where the last inequality holds near and after
decoupling. If the opposite case holds, $\tau_{\rm therm}>\tau_{\rm
dyn}$, the protons could keep the velocity dispersion they acquire
through collisions with neutrons ({\it i.e.} become ``hot'').  

\subsection{Collisional re-thermalization mechanisms}

There are two energy loss mechanisms for protons with a velocity relative
to the background plasma. The first is proton-electron scattering. Bethe's 
formula \cite{bethe} gives this rate as:
\begin{equation}
\label{tautherm1}
\left({\frac{dE}{dt}}\right)_{{\rm p}e} \approx 
\left({7.6\times 10^{-18}}{\rm GeV}{\rm \, s^{-1}}\right) 
{\frac{n_e}{v}} \left[ \ln\left({\frac{2 \gamma ^2 v^4}
{n_e}}\right) + 74.1 \right]
\end{equation}
Here $n_e$ is the electron number density in units of $\rm cm^{-3}$ and
$v$ and $\gamma$ refer to the velocity and Lorentz factors respectively,
 of the proton
relative to the plasma.

The relations derived above imply that decoupling occurs at a
temperature 
$ T_{\rm decouple} \approx { T_{\rm D}} ({s_5}/{(v_{\rm rel}
\sigma_{10})_{\rm dec} Y_e \tau_{5}})^{1/3}$, with 
${T_{\rm D}}=0.005\, {\rm MeV}$. This clearly implies that decoupling
occurs after $e^{\pm}$ annihilation. Noting that the ionization electron
number density is $n_e \sim 10^{27} T^3_{\rm
MeV} s_5 ^{-1} Y_e {\rm cm}^{-3}$, and noting that typical kinetic energies of
protons relative to the plasma are of the order of a few GeV, we see that
the
thermalization timescale via proton/electron scattering is 
$
\tau_{\rm therm,{\rm p}e}\approx 8\cdot 10^{-5}\ {\rm s} 
\left({ T_{\rm D}}/{T} \right)^3 s_5/Y_e.
$

Ref.~\cite{derishev} made the observation that a
pion-induced electromagnetic cascade may lead to an increase in the
number density of electrons and positrons in the fireball. The
greatest this increase could be occurs when all of the pion mass
(minus neutrino losses) goes into $e^+/e^-$ pairs and when, in addition,
the processes creating $e^+/e^-$ pairs are sufficiently rapid that
they come into equilibrium with electron/positron pair
annihilation. 

By considering this limiting case, one finds that a 
pion-induced electromagnetic cascade can delay the onset of a dispersion in
the proton component ({\it i.e.} make $\tau_{\rm therm,{\rm p}e}\left(T=T_{\rm
decouple}/4\right)<\tau_{\rm dyn}$) only for very small electron
fractions, $Y_e<0.01$. (The number 4 appearing in parentheses here
arises from requiring that a large center of mass energy collision is
needed to drive a proton dispersion and is explained in more detail
below). Therefore, a pion induced electromagnetic cascade has little
or no leverage in delaying the development of a dispersion in the
protons for moderate electron fractions.

The second energy loss mechanism for protons arises because, when the protons
acquire a dispersion, the electrons must rearrange themselves 
in order to preserve
charge neutrality. This rearrangement, on its own, has a negligible effect
on the proton dispersion because of the smallness of the ratio of the 
electron to proton mass. 
However, Compton scattering following this rearrangement
can be a source of energy loss for the protons. 
In other words, runaway protons may be viewed as being tightly paired with
electrons via the coulomb force, and the proton loses energy not only
at the rate given in Eq. \ref{tautherm1}, 
but in addition at the rate at which the electron 
to which it is coupled loses energy to the background photons
via Thomson/Compton drag:

\begin{equation}
\label{tautherm2}
\left({\frac{dE}{dt}}\right)_{{\rm p}e\gamma}
\approx \sigma_{\rm T}U_{\gamma} \approx \left( 10^6 {\rm GeV}{\rm
s^{-1}}\right)
\left(\frac{T}{ T_{\rm D}}\right)^4, 
\end{equation}
with $U_{\gamma}$ the photon energy density, giving,
$
\tau_{{\rm therm},{{\rm p}e\gamma}}
\approx 10^{-6} {\rm s} 
\left({{ T_{\rm D}}}/{T}\right)^4. 
$
The total thermalization timescale is given by 
\begin{equation}\tau_{\rm therm}^{-1}=
\tau_{{\rm therm},{{\rm p}e\gamma}}^{-1}+\tau_{{\rm therm},{{\rm p}e}}^{-1}.
\end{equation}

A hot proton component arises, then, if neutron decoupling occurs, and at the
end of the fireball's acceleration phase 
$\tau_{\rm therm}>\tau_{\rm dyn}$. Applying the
simple picture given above for the evolution of the fireball gives the second
condition for a hot proton component:
\begin{equation}
s_5^4 > (400+Y_e)\tau_{5}.
\end{equation}

In fact, the fireball is still accelerating for fireball radius values
greater than $\eta R_0$
(\cite{kobayashiandpiran}, and see below).  This implies that the temperature 
during the
acceleration phase becomes lower than $T_0/\eta$ and, hence,
 that the condition 
for the 
development of a hot proton component is somewhat weaker than that given 
above.

In Fig. 1 we illustrate the decrease in degree of thermalization with 
increasing initial $\tau_{\rm therm}/
\tau_{\rm dyn}$ for a proton moving relative to a background plasma and
losing energy via the  the processes
described in Eqs. \ref{tautherm1} and \ref{tautherm2}.

\begin{figure}
\includegraphics[width=3in]{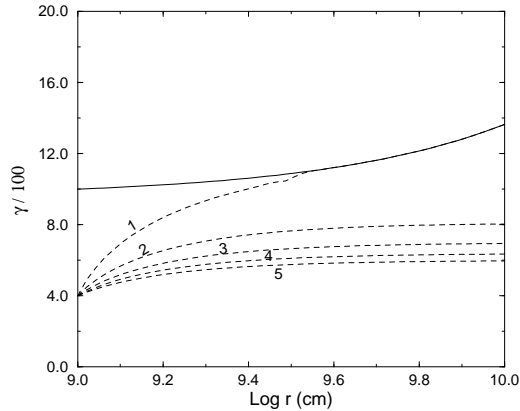}
\caption{\small 
Illustration of the evolution of Lorentz factor of a test proton 
in a background plasma. The solid line is for the Lorentz factor of the 
background 
plasma and the dashed lines are for the Lorentz factors of the test protons. 
Different dashed lines
correspond to different values of $\tau_{\rm therm}/\tau_{\rm dyn}$ (i.e. to 
different plasma temperatures) at 
$10^9$cm. The different
values of $\tau_{\rm therm}/\tau_{\rm dyn}$ are given next to the curves.}
\end{figure}
\subsection{Collisionless considerations}   
In addition to the requirement of charge neutrality leading to proton
energy loss via electron Compton scattering, there are other collisionless
processes that may affect the dynamics of the protons.
Indeed, the proton and electron plasma frequencies are
comparable to or larger than the collision frequencies after decoupling, 
\begin{equation}
\omega_{\rm pi}\sim 10^{11} s^{-1}(Y_es_5^{-1}(T/T_{\rm D})^3)^{1/2}\sim (1/43)
\omega_{\rm pe},\end{equation} 
with $\omega_{\rm pi}$ and $\omega_{\rm pe}$ the proton and electron
plasma frequencies. The post decoupling proton dispersed fireball
may provide fertile and novel ground for the study of plasma instabilities.
In particular, the protons may be subject to an analog of the two-stream 
instability because of collisionally produced bumps in the proton distribution
function at the bulk plasma and near the mean neutron component 
velocities. 

The plasma may also be subject to velocity space anisotropy-driven 
electromagnetic instabilities as the transverse velocity freezes
out during the expansion of the fireball. Relevant for the 
present work, it has been shown that magnetic 
trapping can saturate these instabilities at substantial ion 
anisotropies \cite{saturation},
and that they are moderately stabilized by relativistic effects 
\cite{yoon0,yoon}. A detailed study of the saturation level of the proton 
anisotropy in this system would be interesting because lower saturation levels
imply less dispersion reduction due to adiabatic cooling of the
tranverse momenta. In the presence of a magnetic field,
either generated by the central engine or created through instabilities,
other instabilities, involving, for example, ion-cyclotron waves, may operate.
 
Instabilities may play an 
important role in the setting of the proton distribution function. They are not
expected to alter our result of a dispersion in the proton component, because
for the closely related system of counter-streaming ion beams they are well
studied and a substantial reduction in an initial dispersion is not seen
\cite{saturation,forslund,krall}.
These instabilities have also been studied
in the regime where collisions are important, and in this case 
counter-streaming ion beams take, as expected, a few collision timescales
to slow \cite{berger}. However, studies 
of the coupling between plasma instabilities
and the hydrodynamic evolution of the protons in the peculiar high photon
entropy/small dynamic timescale environment characteristic of GRB fireballs
have apparently not been done, and may 
give interesting and unexpected results. The fact that plasma instabilities
play a role during the acceleration phase of some relativistic 
fireballs is quite interesting, because the previous picture was that 
collisional processes are sufficient to describe the early evolution in the
absence of central engine-generated magnetic fields. 

A diagram illustrating
the various influences on the proton distribution function, 
including the trend toward a smoother distribution function resulting from
plasma instabilities, is shown in Fig. 2.

\begin{figure}
\includegraphics[width=3in]{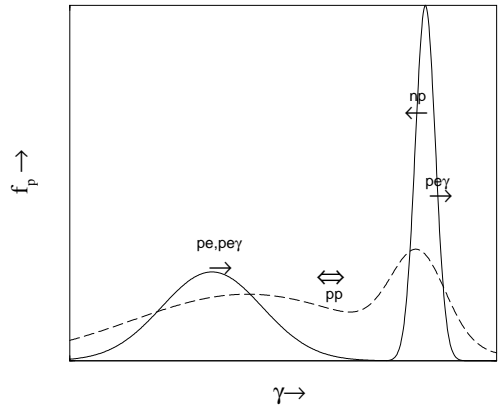}
\caption{\small 
A schematic picture of the various influences on the proton Lorentz 
factor distribution 
function during the acceleration of the fireball. The solid line represents
a collisionally produced 
proton distribution function, 
while the dashed line represents the tendency of instabilities to 
smooth the distribution as collisionless processes liberate 
free energy. In this figure, pp and np represent proton-proton and 
neutron-proton collisions, while pe and pe$\gamma$ represent the
processes in Eqs. \ref{tautherm1} and \ref{tautherm2}, respectively.  
}
\end{figure}

\subsection{Dispersion in the proton component: analytic and numerical
estimates}

When neutron decoupling occurs, and is pronounced, 
a rough estimate gives the dispersion
in proton Lorentz factor as 
$\langle{\gamma_{\rm p,{\rm final}}}\rangle-\gamma_{{\rm
n,final}}.$
(A somewhat more precise measure of dispersion will be given below.)
For example, consider a fireball described by $Y_e = 0.2,\tau_{5}=2.7$
and suppose that $T_0=3$ and $s_5=4$ (implying $\eta \sim
1000$). We find $\gamma_{n,\rm final}\sim 340$ and $\bar{\gamma}_{{\rm p,
final}} \sim 3.4 \times 10^{3}$ implying a dispersion in the proton
component of $10^{3} \sim \eta$. 

We have defined the dispersion in terms
of the central engine rest frame. For some purposes it is more useful
to express the dispersion in terms of the mean proton rest frame. A simple 
approximate measure of the dispersion in this frame is 
\begin{equation} 
\frac{1}{2}
\left(\frac{ \langle{\gamma_{\rm p,{\rm final}}}\rangle }
{\gamma_{\rm n, {\rm final}}}\right)^{1/2}
 \left(
\gamma_{\rm n, {\rm final}}/
{\langle{\gamma_{\rm p,{\rm final}}}\rangle}+1\right),
\end{equation} 

which is at most of order 
a few.

A more precise determination of the dispersion in the protons requires
a transport calculation describing the evolution of the fireball. A
simple single angular zone, hydrodynamically consistent, steady state
relativistic wind transport calculation has been performed. Neutrino energy 
loss is estimated by assuming that following an inelastic collision the 
available kinetic energy is shared equally between two baryons and a pion.
The pion energy is then assumed lost from the system.
 In the calculation, protons suffering large energy
($\gamma_{\rm rel}>2$) collisions with neutrons are assumed to
decouple from the plasma if $\tau_{\rm therm}> \alpha \tau_{\rm dyn}$,
where $\alpha$ is taken as a free parameter. 
 In the limit where
neutrons and protons are coupled to the photon/$e^{\pm}$ plasma our
calculation reduces to that done in the pioneering study by
Paczy\'nski~\cite{pac} on ultra-relativistic winds from compact
sources.

In Fig. 3 we present the results of a calculation for the parameters
listed above: $\tau_{5}=2.7$, $T_0=3$, $s_5=4$ and for the cases where
$Y_e=0.1,0.2,0.3,$ and $0.5$. Results for different values of $Y_e$
have been presented to illustrate that the lower the electron
fraction, the more pronounced the effects of collisions between
neutrons and protons.

\begin{figure}
\includegraphics[width=3in]{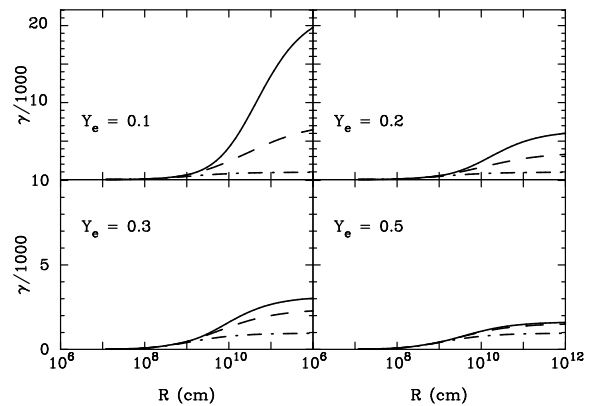}
\caption{\small Evolution of proton Lorentz factor in a steady state
relativistic wind characterized by the parameters given in the text,
for different cases of electron fraction, $Y_e$. The dot dashed line 
is for
case (i), where  baryons are assumed coupled to the plasma; the dashed line is
for case (ii), where neutron velocities 
are described by a velocity distribution function and where neutrons
are allowed to decouple while protons are assumed frozen into the
plasma; and the solid line is for case (iii), where neutrons and protons 
are both
allowed to decouple.  Note the change in scale between the upper and 
lower figures.}
\end{figure}

To illustrate the effects discussed here we present results for the
case where (i) protons and neutrons are assumed coupled to and
thermalized with the plasma; (ii) a distribution function is used to
describe the neutrons but the protons are assumed coupled to and
thermalized with the $e^{\pm}/$photon plasma (the agreement between 
the final proton Lorentz factors in this case and the simple estimate
given by Eq. \ref{gammabarp} is good); and (iii) a velocity distribution
function is used to describe the neutron velocities, and protons are assumed
to decouple from the plasma once they have undergone a collision with
a neutron where $\gamma_{\rm rel}>2$. For part (iii) of the calculation,
the parameter $\alpha$ was taken to be 5. For the example parameters we
have chosen the condition $\tau_{\rm therm} > 5 \tau_{\rm dyn}$ is always
satisfied at the stage in the fireball's evolution when  
$\gamma_{\rm rel}>2$ collisions occur, and so the first condition 
is somewhat superfluous here.

Perhaps the most striking feature of this calculation is 
the increase in Lorentz factor of the highest energy protons as the protons
acquire a dispersion.  Because energy conservation implies that the
mean Lorentz factor of the protons is unchanged between cases (ii) and
(iii) (modulo changes in the net energy carried by the neutrons and
pions due to the protons acquiring a velocity dispersion) the
difference in final Lorentz factors between cases (ii) and (iii)
provides a measure of the proton dispersion. Note that because our
calculation does not allow for a proton to interact with the
expanding plasma once it has undergone a high energy collision with a
neutron and once $\tau_{\rm therm}>5 \tau_{\rm dyn}$ our calculation
may somewhat overestimate the proton dispersion.

In Fig. 4 we display the evolution of a measure of dispersion in the
proton component of the fireball ($\Delta\gamma/\gamma_{{\rm p,final}}$).
  In this figure $\Delta\gamma$ is defined to be the
difference in proton Lorentz factors between models (ii) and (iii) for
the case $Y_e=0.2$ shown in Fig. 1.

\begin{figure}
\includegraphics[width=3in]{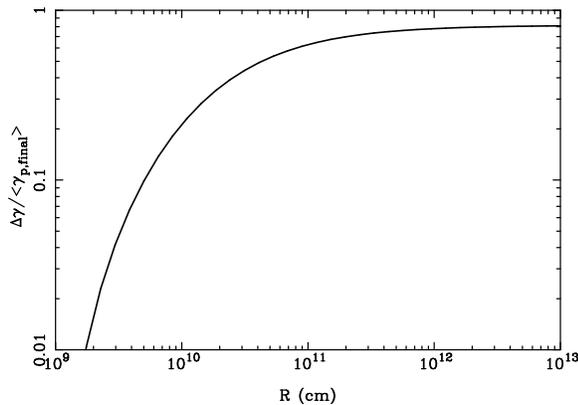}
\caption{\small Evolution of $\Delta\gamma/
\langle{\gamma_{p,{\rm final}}}\rangle$
with radius for the case $Y_e = 0.2$ and the parameters given in the
text.}
\end{figure}

\section{Conclusions}
We have pointed out and made estimates of an overlooked (and possibly
generic) feature of the baryon (neutron/proton) flow in GRB fireballs:
the acquisition of a large dispersion in Lorentz factor of the proton
component subsequent to neutron decoupling. 
This represents a qualitatively new feature of GRB fireballs
because the previous picture of the evolution of the proton component
 held that the protons were thermal, even when neutron decoupling
occurs. A dispersion in the proton component also implies that plasma
instabilities may play an important role and suggests new
studies of the role of collisionless processes during the acceleration 
and later shocking phase of the fireball's evolution. Because the effect
we have described is principally sensitive to the neutron to proton ratio
in the fireball, this work represents a new connection between the central
engine weak physics that sets $Y_e$ and GRB fireball dynamics. 
Because shocks involving protons are thought to give rise
to the photon signal characteristic of GRBs, a proton dispersion
also has implications for
the details and efficiency of photon production.
A direct consequence for the
charged particle dynamics of neutron decoupling
is interesting because the signal from the decay of
pions created in inelastic collisions during neutron decoupling is
expected to be weak.   Detector event rates for the neutrinos from pion
decay are estimated at a few per year \cite{bahmes}. 
Direct evidence of neutron decoupling in a GRB event would provide
information about the progenitor fireball parameters and perhaps give a clue
about the properties of the central engine. 

This work was partially supported by NSF Grant PHY98-00980 at UCSD, an
IGPP mini-grant at UCSD, and a NASA GSRP for KA.  We are indebted to
Neal Dalal, 
C.Y.~Cardall, Tom O'Neil, Pat 
Diamond, and Dan Dubin for useful insights.

\end{document}